\documentclass[12pt]{article}
\usepackage[dvips]{color}
\usepackage{epsfig}
\usepackage{amsmath}
\usepackage{graphicx}
\newcommand{\be}{\begin{equation}}
\newcommand{\ee}{\end{equation}}
\begin{document}

\title{\bf Calculating the jet-quenching parameter in STU background}
\author{K. Bitaghsir Fadafan $^{a}$\thanks{Email: bitaghsir@shahroodut.ac.ir}\hspace{1mm}
B. Pourhassan $^{b}$ \thanks{Email: b.pourhassan@umz.ac.ir}
and J. Sadeghi $^{b,c}$ \thanks{Email: pouriya@ipm.ir}\\\\
$^{a}$ \emph{Physics Department, Shahrood University of Technology}\\
{\emph{P.O.Box 3619995161, Shahrood, Iran}}\\
$^b$ \emph{Physics Department, Mazandaran University}\\
{\emph{P.O.Box 47416-95447, Babolsar, Iran}}\\
$^{c}$ \emph{Institute for Studies in Theoretical Physics and
Mathematics (IPM)}\\
{\emph{P.O.Box 19395-5531, Tehran, Iran}}} \maketitle

\begin{abstract}
\noindent In this paper we use the AdS/CFT correspondence to compute
the jet-quenching parameter in a $\mathcal{N}=2$ thermal plasma. We
consider the general three-charge black hole and discuss some
special cases. We add a constant electric field to the background
and find the effect of the electric field on the jet-quenching
parameter. Also we include higher derivative terms and obtain the
first-order
correction for the jet-quenching parameter.
\end{abstract}
\newpage
\tableofcontents
\newpage
\section{Introduction}
The AdS/CFT correspondence, proposed by Maldacena [1] and developed
by Witten [2] and Gubser,  et al. [3], is a relation between a
supergravity theory in the $d+1$-dimensional anti de Sitter (AdS)
space and a conformal field theory (CFT) in the $d$-dimensional
boundary of AdS space. The famous example of AdS/CFT correspondence
is the relation between type IIB string theory in $AdS_{5}\times
S^{5}$ space and $\mathcal{N}=4$ super Yang-Mills gauge theory on
the 4-dimensional boundary of $AdS_{5}$ space. This duality gives us
useful tools for studying QCD. An important problem in QCD, which is
also interesting to experiment in the LHC and RHIC [4-7], is
considering a moving heavy quark through the hot plasma [8]. This
problem and related topics have been already studied by using the
AdS/CFT correspondence [9-18]. There are also other thermodynamical
quantities such as the free energy, the energy density, the heat
capacity and the speed of sound which
can be calculated and compared with RHIC data [4-7].\\
Another important property of the strongly-coupled plasma at RHIC is
the ratio of the shear viscosity to the entropy density. The
universality of the ratio of shear viscosity $\eta$ to entropy
density $s$ [19, 20, 21, 22] for all gauge theories with Einstein
gravity dual raised the tantalizing prospect of a connection between
string theory and RHIC. The results were obtained for a class of
gauge theories whose holographic duals are dictated by classical
Einstein gravity. However, in the gravity side one can study higher
curvature AdS models and show that the conjectured lower bound on
the
$\frac{\eta}{s}$ can be violated [23].\\
Recently there have been many searches for less supersymmetric model
than $\mathcal{N}=4$, for example $\mathcal{N}=2$ supergravity
theory [24-31]. It is important to note that solutions of
$\mathcal{N}=2$ supergravity may be solutions of a supergravity
theory with more supersymmetry such as $\mathcal{N}=4$ and
$\mathcal{N}=8$. Also the AdS supergravity can obtained by gauging
the $U(1)$ group in $\mathcal{N}=2$ supersymmetric algebra. We have
Already studied the problem of the drag force of a moving quark and
also quark-antiquark pair through the $\mathcal{N}=2$ thermal plasma
[28, 29, 30]. In the Ref. [28] we considered the same background
with present paper and studied the drag force of moving quark
through thermal plasma, we calculated the quasi-normal modes
corresponding to the static string. Also we obtained the effect of
the constant electromagnetic field on the energy-momentum current
densities. Furthermore we considered the effect of higher derivative
corrections. In the Ref. [28] we found that the problem of the drag
force in $\mathcal{N}=2$ supergravity theory at near-extremal limit
is corresponding to the
problem of the drag force in $\mathcal{N}=4$ SYM theory.\\
In this paper we are going to obtain the jet-quenching parameter for
the $\mathcal{N}=2$ thermal plasma. Jet-quenching is a property of
the quark-gluon plasma (QGP) and our knowledge about this parameter
increases our understanding about the QGP. In that case the
jet-quenching parameter obtained by calculating the expectation
value of a closed light-like Wilson loop and using the dipole
approximation [32]. Therefore one needs to describe the system in
light-cone coordinates. In the description of the AdS/CFT
correspondence the endpoint of both fundamental and Dirichlet
strings under influence of non-zero NS NS B-field background
corresponds to the moving quark with a constant electromagnetic
field. Therefore it is interesting to add a constant $B$ field [33]
to the system and find the effect of constant electric field on the
jet-quenching parameter. In order to calculate this parameter in QCD
one needs to use perturbation theory. But by using AdS/CFT
correspondence the jet-quenching parameter calculated in
non-perturbative quantum field theory. This calculations were
already performed in the $\mathcal{N}=4$ super Yang-Mills thermal
plasma [34-39]. Also the effect of higher derivative corrections
such as Gauss-Bonnet on the drag force and the jet-quenching
parameter has been studied [39, 40]. It is shown that the jet
quenching parameter is enhanced due to the Gauss-Bonnet corrections
with positive $\lambda_{GB}$, while $\hat{q}$ decreases
with negative $\lambda_{GB}$.\\
This paper is organized as follows, in section 2 we review the
$D=5$, $\mathcal{N}=2$ supergravity known as STU model. Then in
section 3, we calculate the jet-quenching parameter in the
$\mathcal{N}=2$ background. In section 4 we add a constant electric
field to the system and discuss the effect of it on the
jet-quenching parameter. Finally in section 5 we summarize our
results.
\section{STU background}
In this section we introduce the three-charge non-extremal black
hole solution in $\mathcal{N}=2$ supergravity which is called the
STU model [27, 41]. In this model, there are three real scalar
fields $X^{i}$ which is corresponding to three black hole charges.
These fields satisfy in a condition as $X^{1}X^{2}X^{3}=1$. In this
paper we discuss the special cases of the (I) one-charge black hole,
$q_{1}=q, q_{2}=q_{3}=0$, (II) two-charge black hole,
$q_{1}=q_{2}=q, q_{3}=0$, and (III) three-charge black hole,
$q_{1}=q_{2}=q_{3}=q$. In the general case, the STU model is
described by the following solution [41],
\begin{equation}\label{s1}
ds^{2}=-\frac{f}{\mathcal{H}^{\frac{2}{3}}}dt^{2}+\mathcal{H}^{\frac{1}{3}}(\frac{r^{2}}{R^{2}}d\vec{x}^{2}+\frac{dr^{2}}{f}),
\end{equation}
where,
\begin{eqnarray}\label{s2}
f&=&1-\frac{\eta}{r^{2}}+\frac{r^{2}}{R^{2}}\mathcal{H} ,\nonumber\\
\mathcal{H}&=&\prod_{i=1}^{3}(1+\frac{q_{i}}{r^{2}}).
\end{eqnarray}
In the equation (1) the parameter $r$ is the radial coordinate along
the black hole, so the boundary of the AdS space located on the
brane ($r\rightarrow\infty$). In the equation (2) there is an
overall factor $q_{i}=\eta\sinh^{2}\gamma_{i}$, where $\eta$ is
called non-extremality parameter and $\gamma_{i}$ related to the
electric charges of the black hole. Finally we should note that the
above solution includes a three-dimensional sphere, but we assume
that the motion is along $\vec{x}$ directions. The Hawking
temperature of this model is given by,
\begin{equation}\label{s3}
T=\frac{2+\sum_{i=1}^{3}k_{i}-\prod_{i=1}^{3}k_{i}}{2\prod_{i=1}^{3}(1+k_{i})^{\frac{1}{2}}}\frac{r_{h}}{\pi
R^{2}},
\end{equation}
where $k_{i}\equiv\frac{q_{i}}{r_{h}^{2}}$ and the constant $R$
denotes the curvature of the AdS space-time. The radius $r_{h}$
denotes the event horizon of the black hole. Previous studies of the
jet-quenching parameter focused on the near-extremal black hole in
the $\mathcal{N}=4$ super Yang-Mills theory. Here, we can see that
at the $\eta\rightarrow0$ limit, where there is near-extremal black
hole, the equation (3) reduces to the Hawking temperature in
$\mathcal{N}=4$ super Yang-Mills theory, where $T=\frac{r_{h}}{\pi
R^{2}}$.
Also the zero temperature limit obtained by putting $r_{h}^{2}=\frac{q_{i}}{2}$ or $k_{i}=2$ in the equation (3).\\
As explained in the introduction, in order to obtain the
jet-quenching parameter one needs to rewrite the metric (1) in the
light-cone coordinates. Therefore one can introduce new coordinates
as $x^{\pm}=\frac{t\pm x^{1}}{\sqrt{2}}$ and rewrite the metric (1)
in the following form,
\begin{eqnarray}\label{s4}
ds^{2}&=&\frac{1}{2}(\frac{\mathcal{H}^{\frac{1}{3}}r^{2}}{R^{2}}-\frac{f}{\mathcal{H}^{\frac{2}{3}}})
\left((dx^{+})^{2}+(dx^{-})^{2}\right)
-(\frac{\mathcal{H}^{\frac{1}{3}}r^{2}}{R^{2}}+\frac{f}{\mathcal{H}^{\frac{2}{3}}})dx^{+}dx^{-}\nonumber\\
&+&\mathcal{H}^{\frac{1}{3}}\left(\frac{r^{2}}{R^{2}}(dx_{2}^{2}+dx_{3}^{2})+\frac{dr^{2}}{f}\right).
\end{eqnarray}
We need to obtain the lagrangian density by using the above metric
and put in the following Nambu-Goto action,
\begin{equation}\label{s5}
S=-\frac{1}{2\pi\alpha^{\prime}}\int{d\tau d\sigma \sqrt{g}}.
\end{equation}
In the next section we obtain the jet-quenching parameter in the
above background.
\section{Jet-quenching parameter}
We begin with the general relation for the jet-quenching parameter
[34],
\begin{equation}\label{s6}
\hat{q}\equiv8\sqrt{2}\frac{S_{I}}{L^{-}L^{2}},
\end{equation}
where $S_{I}=S-S_{0}$. Therefore, calculating the jet-quenching
parameter reduces to obtain actions $S$ and $S_{0}$. At the first,
one should consider an open string whose endpoints lie on the brane.
In the light-cone coordinates, the string may be described by
$r(\tau, \sigma)$. We use the static gauge where $\tau=x^{-}$ and
$\sigma=x^{2}\equiv y$, and all other coordinates considered as
constants. In that case $-\frac{L}{2}\leq y \leq\frac{L}{2}$, and
$L^{-}\leq x^{-} \leq0$, and because of $L^{-}\gg L$ one can assume
that the world-sheet is invariant along the $x^{-}$ direction.
Therefore the string may described by the $r(y)$, so the boundary
condition is $r(\pm\frac{L}{2})=\infty$. In this configuration, the
induced metric on the string fulfills obtained as the following,
\begin{equation}\label{s7}
2g=(\frac{\mathcal{H}^{\frac{2}{3}}r^{2}}{R^{2}}-\frac{f}{\mathcal{H}^{\frac{1}{3}}})
(\frac{r^{2}}{R^{2}}+\frac{{r^{\prime}}^{2}}{f}).
\end{equation}
Since the equation (7) is $x^{-}$ dependent, one can integrate over
$x^{-}$ easily and then the Nambu-Goto action is given by,
\begin{equation}\label{s8}
S=\frac{\sqrt{2}L^{-}}{2\pi\alpha^{\prime}}\int_{0}^{\frac{L}{2}}{dy
\sqrt{(\frac{\mathcal{H}^{\frac{2}{3}}r^{2}}{R^{2}}-\frac{f}{\mathcal{H}^{\frac{1}{3}}})
(\frac{r^{2}}{R^{2}}+\frac{1}{f}{r^{\prime}}^{2})}}.
\end{equation}
We can remove the $r^{\prime}$ by using equation of motion. In that
case, since the lagrangian density is time-dependent, one can write,
\begin{equation}\label{s9}
{\mathcal{H}}=\frac{\partial{\mathcal{L}}}{\partial
r^{\prime}}r^{\prime}-{\mathcal{L}}=Const.\equiv E.
\end{equation}
Therefore we can obtain the following relation,
\begin{equation}\label{s10}
r^{\prime2}=\frac{fr^{2}}{R^{2}E^{2}}\left[\frac{\mathcal{H}^{\frac{1}{3}}}{2R^{2}}
(\frac{\mathcal{H}^{\frac{1}{3}}r^{2}}{R^{2}}-\frac{f}{\mathcal{H}^{\frac{2}{3}}})r^{2}-E^{2}\right].
\end{equation}
The equation (10) has two poles where $r^{\prime}=0$. The main pole
exist at the horizon, so it is clear that the equation (10) has a
zero at the horizon where $f=0$. In this case the string comes from
infinity ($r(\frac{L}{2})=\infty$) and touches the horizon and
return to infinity ($r(-\frac{L}{2})=\infty$). The second pole of
the equation (10) obtained by
$\frac{fr^{2}}{\mathcal{H}^{\frac{1}{3}}R^{2}}-\frac{\mathcal{H}^{\frac{2}{3}}r^{4}}{R^{4}}+2E^{2}=0$.
Ref. [42] show is that the string world sheet has one end at a
Wilson line at the boundary with $Im[t]=0$ and the other end at a
Wilson line the boundary with $Im[t]=-i\epsilon$. The only way that
the string worldsheet linking these two Wilson lines can meet is if
the string worldsheet hangs down to the horizon. Therefore the only
physical situation is the first case where the string touches the
horizon. Also in our case, drawing the $r^{\prime2}$ in terms of $r$ tell us that the turning point of string should be $r_{h}$. \\
By using equation (10) in (8) and also new definition of
$B\equiv\frac{1}{E^{2}}$ one can rewrite the Nambu-Goto action in
the following form,
\begin{equation}\label{s11}
S=\frac{L^{-}\sqrt{B}}{2\pi\alpha^{\prime}}\int_{r_{h}}^{\infty}{dr
\frac{r(\frac{\mathcal{H}^{\frac{2}{3}}r^{2}}{R^{2}}-\frac{f}{\mathcal{H}^{\frac{1}{3}}})}
{\sqrt{\frac{\mathcal{H}^{\frac{1}{3}}}{2}(\frac{\mathcal{H}^{\frac{1}{3}}
r^{2}}{R^{2}}-\frac{f}{\mathcal{H}^{\frac{2}{3}}})Bfr^{2}-fR^{2}}}}.
\end{equation}
For the low energy limit ($E\rightarrow0$) we expand the equation
(11) to leading order in $\frac{1}{B}$. This is reasonable since the
determination of $\hat{q}$ demands the study of the small separation
limit of $L$. Then at the first order of $\frac{1}{B}$ one can
obtain,
\begin{equation}\label{s12}
S=\frac{L^{-}}{2\pi\alpha^{\prime}}\int_{r_{h}}^{\infty}{dr
\sqrt{\frac{2\mathcal{H}^{\frac{1}{3}}}{f}(\frac{\mathcal{H}^{\frac{1}{3}}r^{2}}{R^{2}}-\frac{f}{\mathcal{H}^{\frac{2}{3}}}
)}
\left[1+\frac{R^{2}}{(\frac{\mathcal{H}^{\frac{2}{3}}r^{2}}{R^{2}}-\frac{f}{\mathcal{H}^{\frac{1}{3}}})Br^{2}}\right]}.
\end{equation}
Now, we are going to extract action $S_{0}$ which can be interpreted
as the self energy of the isolated quark and the isolated antiquark.
In that case by using the following relation [38],
\begin{equation}\label{s13}
S_{0}=\frac{L^{-}}{\pi\alpha^{\prime}}\int_{r_{h}}^{\infty}{dr
\sqrt{G_{--}G_{rr}}},
\end{equation}
one can find,
\begin{equation}\label{s14}
S_{0}=\frac{L^{-}}{2\pi\alpha^{\prime}}\int_{r_{h}}^{\infty}{dr
\sqrt{\frac{2\mathcal{H}^{\frac{1}{3}}}{f}(\frac{\mathcal{H}^{\frac{1}{3}}r^{2}}{R^{2}}-\frac{f}{\mathcal{H}^{\frac{2}{3}}}
)}}.
\end{equation}
Then one can obtain easily,
\begin{equation}\label{s15}
S_{I}=\frac{1}{\sqrt{B}}\frac{L^{-}}{2\pi\alpha^{\prime}}\int_{r_{h}}^{\infty}{dr
\sqrt{\frac{2R^{4}}{(\frac{\mathcal{H}^{\frac{2}{3}}r^{2}}{R^{2}}-\frac{f}{\mathcal{H}^{\frac{1}{3}}})Bfr^{4}}}}.
\end{equation}
On the other hand, one can integrate equation (10) and obtain the
following relation for infinitesimal $\frac{1}{B}$,
\begin{equation}\label{s16}
\frac{L}{2}=R^{2}\int_{r_{h}}^{\infty}{dr
\frac{1}{\sqrt{\frac{B}{2}(\frac{\mathcal{H}^{\frac{2}{3}}r^{2}}{R^{2}}-\frac{f}{\mathcal{H}^{\frac{1}{3}}})fr^{4}}}}.
\end{equation}
Therefore, by using relations (6), (15) and (16) we can specify the
jet-quenching as the following,
\begin{equation}\label{s17}
\hat{q}=\frac{(I(q))^{-1}}{\pi\alpha^{\prime}}.
\end{equation}
where
\begin{equation}\label{s18}
I(q)=R^{2}\int_{r_{h}}^{\infty}{\frac{dr}
{\sqrt{(\frac{\mathcal{H}^{\frac{2}{3}}r^{2}}{R^{2}}-\frac{f}{\mathcal{H}^{\frac{1}{3}}})fr^{4}}}}.
\end{equation}
Here, it is important to explain horizon structure of the STU
solution. The $f(r)=0$ from the equation (2) reduces to the
following equation [27],
\begin{equation}\label{s19}
r^{6}+{\mathcal{A}}r^{4}-{\mathcal{B}}r^{2}+q_{1}q_{2}q_{3}=0,
\end{equation}
where ${\mathcal{A}}=q_{1}+q_{2}+q_{3}+R^{2}$ and
${\mathcal{B}}=\eta R^{2}-q_{1}q_{2}-q_{2}q_{3}-q_{1}q_{3}$. The
equation (19) has two positive and one negative zero. Two positive
roots interpreted as inner and outer horizon, so the horizon radius
$r_{h}$ denoted the outer horizon.\\
In order to obtain the explicit expression of the jet-quenching
parameter we consider three special cases of one, two and three charged black hole.\\
\subsection{One-charged black hole}
In this case we set $q_{1}=q, q_{2}=q_{3}=0$. So, the integral (18)
reduced to the following expression,
\begin{equation}\label{s20}
I(q_{1})=R^{4}\int_{r_{h}}^{\infty}{\sqrt{\frac{(1+\frac{q}{r^{2}})^{\frac{1}{3}}}
{(r^{2}-\eta)(r^{4}+(q+R^{2})r^{2}-\eta R^{2})}}dr},
\end{equation}
where
\begin{equation}\label{s21}
r_{h}^{2}=\frac{1}{4}(-2q+2\pi^{2}R^{4}T^{2}+2\sqrt{2q\pi^{2}R^{4}T^{2}+\pi^{4}R^{8}T^{4}}).
\end{equation}
The relation (21) obtained from the Hawking temperature (3). On the
other hand from the equation (19) one can obtain,
\begin{equation}\label{s22}
r_{h}^{2}=\frac{q+R^{2}}{2}\left[-1+\sqrt{1+\frac{4\eta
R^{2}}{(q+R^{2})^{2}}}\right].
\end{equation}
If the condition $\frac{2r_{0}^{2}}{q+R^{2}}\gg1$ satisfied, then
the rescaling $\eta R^{2}\equiv r_{0}^{4}$ implies that
$r_{h}=r_{0}$ and we recover the case of ${\mathcal{N}}=4$ SYM
plasma.\\
Numerically, we draw graph of the jet-quenching parameter in terms
of the black hole charge and the temperature in the Fig. 1 and Fig.
2 respectively. These plots show that the jet-quenching parameter of
the $\mathcal{N}=2$ theory is larger than the jet-quenching
parameter of the $\mathcal{N}=4$ theory. For example by choosing
$R^{2}=\alpha^{\prime}\sqrt{\lambda}$, $\alpha^{\prime}=0.5$
($\alpha^{\prime}=0.25$), $\lambda=6\pi$, $q=10^6$ and $T=300$ $MeV$
one can obtain, $\hat{q}=42$ ($\hat{q}=18$) $GeV^2/fm$. In that case
the thermodynamical stability let us to choose $q\leq5\times10^6$
for $T=300 MeV$. On the other hand, for the small black hole charge,
by taking $\alpha^{\prime}=0.5$ and $\lambda=6\pi$ one can obtain
$\hat{q}=37.5$ $GeV^2/fm$. It means that the black hole charge
increases the jet-quenching parameter. In order to obtain
$\hat{q}=5$ $GeV^2/fm$ the corresponding temperature of the quark
gluon plasma is $155 MeV$, which is smaller than expected [43].
\subsection{Two-charged black hole}
In this case we set $q_{1}=q_{2}=q, q_{3}=0$. So, the integral (18)
reduced to the following expression,
\begin{equation}\label{s23}
I(q_{1,2})=R^{4}\int_{r_{h}}^{\infty}{\sqrt{\frac{(1+\frac{q}{r^{2}})^{\frac{2}{3}}}
{\rho(r^{4}+(2q+R^{2})r^{2}-\eta R^{2}+q^{2})}}dr},
\end{equation}
where we defined
$\rho\equiv((R^{2}-1)r^{4}+(2qR^{2}-R^{2}-q)r^{2}+R^{2}q^{2}+\eta
R^{2}-q^{2})$, and,
\begin{equation}\label{s24}
r_{h}=\pi R^{2}T.
\end{equation}
The relation (24) obtained from the Hawking temperature (3).
Numerically, we draw graph of the jet-quenching parameter in terms
of the black hole charge and the temperature in the Fig. 1 and Fig.
2 respectively. These plots show that the jet-quenching parameter of
the $\mathcal{N}=2$ theory is larger than the jet-quenching
parameter of the $\mathcal{N}=4$ theory. Also we find that the
jet-quenching parameter of the two-charged black hole is larger than
the jet-quenching parameter of the one-charge black hole. For
example by choosing $R^{2}=\alpha^{\prime}\sqrt{\lambda}$,
$\alpha^{\prime}=0.5$, $\lambda=6\pi$, $q=10^6$ and $T=300$ $MeV$
one can obtain $\hat{q}=49$ $GeV^2/fm$. In that case the
thermodynamical stability let us to choose $q\leq3\times10^6$ for
$T=300 MeV$. If we consider small value of the black hole charge
then find the same value of the jet-quenching parameter as the
previous case, this point illustrated in the Fig. 2. Therefore, in
order to obtain $\hat{q}=5$ $GeV^2/fm$ the corresponding temperature
of the quark gluon plasma is $155 MeV$ for the small black hole
charge.
\subsection{Three-charged black hole}
In this case we set $q_{1}=q_{2}=q_{3}=q$. So, the integral (18)
reduced to the following expression,
\begin{equation}\label{s25}
I(q_{1,2,3})=R^{4}\int_{r_{h}}^{\infty}{\sqrt{\frac{r^{2}(r^{2}+q)}
{\varrho(r^{6}+(R^{3}+3q)r^{4}+(3q^{2}-\eta R^{2})r^{2}+q^{3})}}dr},
\end{equation}
where we defined
$\varrho\equiv((R^{2}-1)r^{6}+(3qR^{2}-R^{2}-3q)r^{4}+(3R^{2}q^{2}+\eta
R^{2}-3q^{2})r^{2}+(R^{2}-1)q^{3})$, and $r_{h}$ obtained from the
Hawking temperature (3). Numerically, we give plot of the
jet-quenching parameter in terms of the black hole charge and the
temperature in the Fig. 1 and Fig. 2 respectively. These plots show
that the jet-quenching parameter of the $\mathcal{N}=2$ theory is
larger than the jet-quenching parameter of the $\mathcal{N}=4$
theory. Also we find that the jet-quenching parameter of the
three-charged black hole is larger than the jet-quenching parameter
of the one-charge and two-charged black holes. For example by
choosing $R^{2}=\alpha^{\prime}\sqrt{\lambda}$,
$\alpha^{\prime}=0.5$, $\lambda=6\pi$, $q=10^6$ and $T=300$ $MeV$
one can obtain $\hat{q}=58$ $GeV^2/fm$. In that case the
thermodynamical stability let us to choose $q\leq2.5\times10^6$ for
$T=300 MeV$. If we consider small value of the black hole charge
then find the same value of the jet-quenching parameter as the
previous cases, this point illustrated in the Fig. 2. Therefore, in
order to obtain $\hat{q}=5$ $GeV^2/fm$ the corresponding temperature
of the quark gluon plasma
is $155 MeV$ for the small black hole charge.\\\\
\begin{figure}[th]
\begin{center}
\includegraphics[scale=.5]{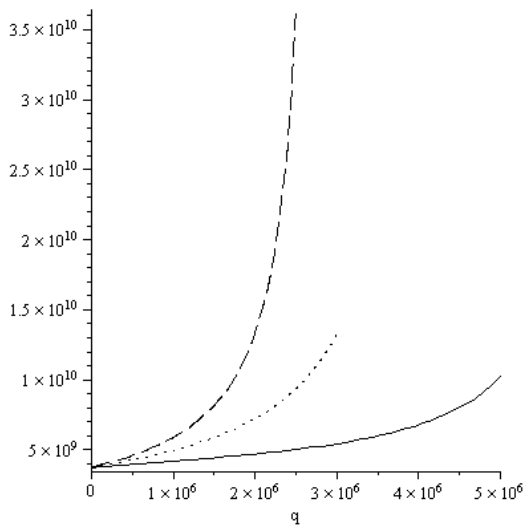}
\caption{Plot of the jet-quenching parameter in terms of the black
hole charge. We fixed our parameter as $\alpha^{\prime}=0.5$,
$\lambda=6\pi$, and $T=300$. The solid line represents the case of
$q_{1}=q, q_{2}=q_{3}=0$. The dotted line represents the case of
$q_{1}=q_{2}=q, q_{3}=0$. The dashed line represents the case of
$q_{1}=q_{2}=q_{3}=q$. It show that increasing the number of black
hole charge increases the value of the jet-quenching parameter.}
\end{center}
\end{figure}

\begin{figure}[th]
\begin{center}
\includegraphics[scale=.5]{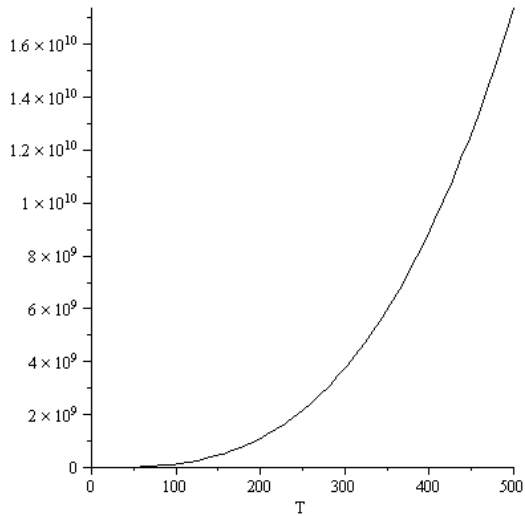}
\caption{Plot of the jet-quenching parameter in terms of the
temperature for small black hole charge. We fixed parameter as
$\alpha^{\prime}=0.5$, $\lambda=6\pi$. In that case three different
cases of one, two, and three-charged black hole have similar
manner.}
\end{center}
\end{figure}

In order to compare our results with the case of $\mathcal{N}=4$ SYM
we also perform the following rescaling [41],
\begin{equation}\label{s26}
r\rightarrow\lambda^{\frac{1}{4}}r, \hspace{5mm}
t\rightarrow\frac{t}{\lambda^{\frac{1}{4}}}, \hspace{5mm}
\eta\rightarrow\lambda\eta, \hspace{5mm}
q\rightarrow\lambda^{\frac{1}{2}}q, \hspace{5mm}
\vec{x}\rightarrow\frac{\vec{x}}{\lambda^{\frac{1}{4}}}.
\end{equation}
Also we set $r_{0}^{4}\equiv\eta R^{2}$, and taking
$\lambda\rightarrow\infty$ yields us to the following result,
\begin{equation}\label{s27}
\hat{q}=\frac{r_{0}^{2}}{\pi\alpha^{'}R^{4}}\left[\int_{r_{h}}^{\infty}{\frac{dr}
{r^{2}\sqrt{\frac{f}{H}}}}\right]^{-1},
\end{equation}
where
\begin{eqnarray}\label{s28}
f&=&H^{3}-\frac{r_{0}^{4}}{r^{4}} ,\nonumber\\
H&=&1+\frac{q}{r^{2}},
\end{eqnarray}
which agree with the results of the Ref. [35], where the
jet-quenching parameter calculated with the chemical potential. The
horizon radius $r_{0}$ obtained for the case of zero-charge black
hole. For the non-zero charge it is clear that the horizon radius
decreases ($r_{h}<r_{0}$). In that case the relation between the
chemical potential and black hole charge may be given by [35, 44],
\begin{equation}\label{s29}
\mu=\frac{\sqrt{2q(q+r_{h}^{2})}}{r_{h}R^{2}}.
\end{equation}
Therefore the $q=0$ limit is equal to $\mu=0$ limit and one can say
that the jet-quenching parameter from the $\mathcal{N}=2$
supergravity theory with zero chemical potential is equal to the
jet-quenching parameter from the $\mathcal{N}=4$ SYM theory.
\section{Effect of constant electric field}
In the description of the AdS/CFT correspondence the endpoint of
both fundamental and Dirichlet strings under influence of non-zero
NS NS B-field background corresponds to the moving quark with a
constant electromagnetic field. In this section we would like to add
a two form $F=edt\wedge dx_{1}$ as a constant electric field to the
line element (1). Antisymmetric field $e$ is the constant electric
field. Now we are going to obtain the effect of the constant
electric field on the jet-quenching parameter. In that case the
Nambu-Goto action is given by the square root of the following
equation,
\begin{equation}\label{s30}
2g=(\frac{\mathcal{H}^{\frac{2}{3}}r^{2}}{R^{2}}-\frac{f}{\mathcal{H}^{\frac{1}{3}}}+e)
(\frac{r^{2}}{R^{2}}+\frac{{r^{\prime}}^{2}}{f}).
\end{equation}
Therefore one can obtain the jet-quenching parameter as the
following,
\begin{equation}\label{s31}
\hat{q}=\frac{(I(q, e))^{-1}}{\pi\alpha^{\prime}},
\end{equation}
where
\begin{equation}\label{s32}
I(q, e)=R^{2}\int_{r_{h}}^{\infty}{\frac{dr}
{\sqrt{(\frac{\mathcal{H}^{\frac{1}{3}}r^{2}}{R^{2}}-\frac{f}{\mathcal{H}^{\frac{2}{3}}}+e)\mathcal{H}^{\frac{1}{3}}fr^{4}}}}.
\end{equation}
where $f$ and $H$ is given by the relation (2). In order to find
effect of the constant electric field on the jet-quenching parameter
we examine above integral for three different cases of one, two and
three-charged black hole. Numerically, and under near boundary
approximation, we draw graph of the jet-quenching parameter in terms
of the constant electric field and find that the constant electric
field increases the value of the jet-quenching parameter. In the
Fig. 3 we draw the jet-quenching parameter in terms of the constant
electric field for the large black hole
charge.\\
\begin{figure}[th]
\begin{center}
\includegraphics[scale=.5]{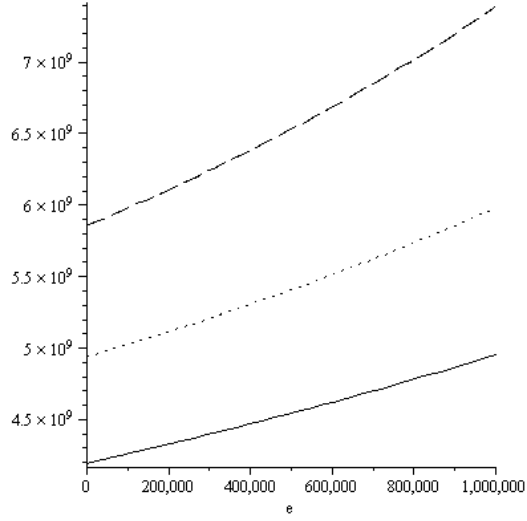}
\caption{Plot of the jet-quenching parameter in terms of the
constant electric field. We fixed our parameter as
$\alpha^{\prime}=0.5$, $\lambda=6\pi$, $q=10^{6}$ and $T=300$. The
solid line represents the case of $q_{1}=q, q_{2}=q_{3}=0$. The
dotted line represents the case of $q_{1}=q_{2}=q, q_{3}=0$. The
dashed line represents the case of $q_{1}=q_{2}=q_{3}=q$. It show
that the jet-quenching parameter increased by the constant electric
field.}
\end{center}
\end{figure}
As before, by rescaling (26) one can obtain,
\begin{equation}\label{s33}
\hat{q}=\frac{1}{\pi\alpha^{\prime}R^{4}}\left[\int_{r_{h}}^{\infty}{\frac{dr}
{r^{4}\sqrt{fH(\frac{r_{0}^{4}}{H^{2}r^{4}}+e)}}}\right]^{-1},
\end{equation}
where $f$ and $H$ is given by the relation (28). It is easy to show
the effect of the constant electric field on the value of the
jet-quenching parameter. The $q=0$ limit of the equation (33)
reduces to the following expression,
\begin{equation}\label{s34}
{\hat{q}}_{0}=\frac{1}{\pi\alpha^{\prime}R^{4}}\left[\int_{r_{0}}^{\infty}{\frac{dr}
{\sqrt{(r^{4}-r_{0}^{4})(e r^{4}+r_{0}^{4})}}}\right]^{-1}.
\end{equation}
It is clear that the case of $e\rightarrow0$ recovers the result of
(27). We find that the effect of constant electric field is
increasing the jet-quenching parameter, so it is in agreement with
increasing the drag force in presence of constant electric field for
ultra relativistic motion [30].
\section{Higher derivative correction}
In this section, in absence of any external field, we would like to
calculate the effect of higher derivative terms on the jet-quenching
parameter. In this way we use results of the Ref. [45], where the
first-order correction of the solution (2) is given by,
\begin{eqnarray}\label{s35}
f&=&1-\frac{\eta}{r^{2}}+\frac{r^{2}}{R^{2}}(1+\frac{q}{r^{2}})^{3}
+\frac{c}{24}\left[\frac{\eta^{2}}{4r^{6}(1+\frac{q}{r^{2}})}-\frac{8q(q+\eta)}{3R^{2}r^{4}}\right] ,\nonumber\\
H&=&1+\frac{q}{r^{2}}-\frac{c}{24}\left[\frac{q(q+\eta)}{3r^{2}(r^{2}+q)^{2}}\right].
\end{eqnarray}
where $c$ is small parameter of the higher derivative correction. We
should note that the solution (35) obtained for the black hole with
three equal charge ($q_{1}=q_{2}=q_{3}=q$). In that case the
jet-quenching parameter obtained as the following expression,
\begin{equation}\label{s36}
\hat{q}=\frac{(I(q, c))^{-1}}{\pi\alpha^{\prime}},
\end{equation}
where
\begin{equation}\label{s37}
I(q, c)=\int_{r_{H}}^{\infty}{\frac{dr}
{\sqrt{(\frac{H^{2}r^{2}}{R^{2}}-\frac{f}{H})fr^{4}}}},
\end{equation}
and $f$ and $H$ are given by the relation (35), also $r_{h}$ is
given by the following equation [45],
\begin{eqnarray}\label{s38}
r_{H}&=&r_{h}\nonumber\\
&+&\frac{cr_{h}}{24}\frac{(1+\frac{q}{r_{h}^2})^{4}(3q^{2}-26q
r_{h}^{2}+3r_{h}^{4})}{24R^{4}(1+\frac{q}{r_{h}^2})
\left[\frac{(1+\frac{q}{r_{h}^2})^{2}(q-2 r_{h}^{2})}{R^{2}}-1\right]}\nonumber\\
&-&\frac{cr_{h}}{24}\frac{2(1+\frac{q}{r_{h}^2})^{2}(13q-3
r_{h}^{2})+3R^{2}}{24R^{2}(1+\frac{q}{r_{h}^2})r_{h}
\left[\frac{(1+\frac{q}{r_{h}^2})^{2}(q-2
r_{h}^{2})}{R^{2}}-1\right]},
\end{eqnarray}

As before one can study near boundary behavior of the jet-quenching
parameter and find that the higher derivative terms include at
$\mathcal{O}(\frac{c}{T^9})$. In that case we find that the higher
derivative terms decrease the value of the jet-quenching parameter.
So, for the fixed parameter such as $\alpha^{\prime}=0.5$,
$\lambda=6\pi$, $T=300$ and small black hole charge, we obtain
$c<0.00021$ to have positive jet-quenching parameter. For example
with the above fixed parameter and $c=0.0001$ one can obtain
$\hat{q}=4.6$ $GeV^2/fm$ which is approximately value of the
jet-quenching parameter of the $\mathcal{N}=4$ SYM theory. In order
to obtain $\hat{q}=5$ $GeV^2/fm$ the corresponding higher derivative
parameter should be $c\approx97\times10^{-4}$ at $T=300 MeV$.\\
Again we use rescaling (26) and take $\lambda\rightarrow\infty$
limit, so we get,
\begin{equation}\label{s38}
\hat{q}=\frac{r_{0}^{2}}{\pi\alpha^{'}R^{4}}\left[\int_{r_{h}}^{\infty}{\sqrt{\frac{H}{f}}\frac{dr}{r^{2}}}\right]^{-1},
\end{equation}
where
\begin{eqnarray}\label{s39}
f&=&(1+\frac{q}{r^{2}})^3-\frac{r_{0}^{4}}{r^{4}}+\frac{c r_{0}^{4}}{24R^{2}r^4}
\left[\frac{r_{0}^{4}}{4r^{2}(r^{2}+q)}-\frac{8q}{3}\right] ,\nonumber\\
H&=&1+\frac{q}{r^{2}}-\frac{c q r_{0}^{4}}{24R^{2}r^{4}(r^{2}+q)},
\end{eqnarray}
and radius $r_{h}$ is the root of the $f=0$ from the relation (40).
The equation (39) may be solved numerically, and explicit expression
of the jet-quenching parameter can be obtained. But it is clear that
the effect of higher derivative correction is to decrease the
jet-quenching parameter. One can check this statement by taking
$q=0$ limit. In this limit the jet-quenching parameter derived as,
\begin{equation}\label{s40}
{\hat{q}}_{0}=\frac{r_{0}^{2}}{\pi\alpha^{\prime}R^{4}}
\left[\int_{r_{h}}^{\infty}{4\sqrt{\frac{6R^{2}r^{4}}{96R^{2}r^{4}(r^{4}-r_{0}^{4})+cr_{0}^{8}}}
\frac{dr}{r}}\right]^{-1},
\end{equation}
where,
\begin{equation}\label{s41}
r_{h}^{4}=\frac{r_{0}^{4}}{2}\left(1+\sqrt{1-\frac{c}{24R^{2}}}\right).
\end{equation}
In that case it is necessary that
$c<24\alpha^{\prime}\sqrt{\lambda}$. Comparing the equation (41)
with the jet-quenching parameter of the $\mathcal{N}=4$ SYM theory
tell us that the effect of $c$ is decreasing the jet-quenching
parameter.
\section{Conclusion}
One of the interesting properties of the strongly-coupled plasma at
RHIC is jet quenching of partons produced with high transverse
momentum. This parameter controls the description of relativistic
partons and it is possible to employ the gauge/gravity duality and
determine this quantity in the finite temperature gauge theories. In
this paper we considered five dimensional $\mathcal{N}=2$ thermal
plasma and calculated the jet-quenching parameter in presence of the
constant electric field, and higher derivative term. We found that
the jet-quenching parameter of the charged black hole of the
$\mathcal{N}=2$ supergravity theory is larger than the
$\mathcal{N}=4$ SYM theory. We examine our solution for three
special cases of one, two and three charged black hole. All cases
yield to the same value of the jet-quenching parameter for the small
black hole charge. However, thermodynamical stability allow to
choose the black hole charge of order $10^{6}$. In that case, by
choosing $\lambda=6\pi$, $\alpha^{\prime}=0.5$ and $T=300 MeV$, we
found $\hat{q}=42, 49$ and $58 GeV^{2}/fm$ for one, two and three
charged black hole respectively. These values of the jet-quenching
parameter are far from experiments of RHIC with above fixed
parameter (experimental data tell us that $5\leq\hat{q}\leq25$).
There is no worry for this statement because the temperature of the
$\mathcal{N}=2$ supergravity theory should given smaller than the
$\mathcal{N}=4$ SYM theory. In that case with the temperature about
$155 MeV$ we obtained the jet-quenching parameter in the
experimental range.\\
We studied the effect of the constant electric field and higher
derivative correction on the jet-quenching parameter. We have shown
that the effect of the constant electric field and the higher
derivative correction is to increase and decrease the jet-quenching
parameter respectively. By analyzing the near boundary behavior we
found the increasing of the jet-quenching parameter due to the
electric field and decreasing of it under the higher derivative is
infinitesimal of order $\mathcal{O}(\frac{e}{T^5})$ and
$\mathcal{O}(\frac{c}{T^9})$ respectively. We found that our results
are agree with the results of the $\mathcal{N}=4$ SYM plasma at
$\eta\rightarrow0$ ($q=0$)
limit or zero chemical potential and rescaling (26).\\

\end{document}